\documentclass{article}
\usepackage{spconf,amsmath,graphicx,array,enumitem,appendix,subcaption}
\usepackage[table,xcdraw]{xcolor}
\usepackage{xcolor}
\title{Accidental Learners: Spoken Language Identification in Multilingual Self-Supervised Models}
\name{Travis M. Bartley$^{\star}$, Fei Jia$^{\dagger}$, Krishna C. Puvvada$^{\dagger}$, Samuel Kriman$^{\dagger}$, Boris Ginsburg$^{\dagger}$\thanks{$^{\star}$Research conducted during internship with NVIDIA.}}
\address{NVIDIA $^{\dagger}$, \\
City University of New York$^{\star}$ 
}

\begin{document}
\maketitle
\begin{abstract}
In this paper, we extend previous self-supervised approaches for language identification by experimenting with Conformer based architecture in a multilingual pre-training paradigm. We find that pre-trained speech models optimally encode language discriminatory information in lower layers. Further, we demonstrate that the embeddings obtained from these layers are significantly robust to classify unseen languages and different acoustic environments without additional training. After fine-tuning a pre-trained Conformer model on the VoxLingua107 dataset, we achieve results similar to current state-of-the-art systems for language identification. More, our model accomplishes this with 5x less parameters. We open-source the model through the NVIDIA NeMo toolkit.
\end{abstract}


\section{Introduction}
\label{sec:intro}
The use of self-supervised (SSL) pre-trained audio models has grown in popularity over recent years, thanks largely to the impressive results demonstrated by the wav2vec architecture~\cite{wav2vec}. Alongside achieving state-of-the-art (SOTA) accuracy for automatic speech recognition (ASR), such models are notable for high performance in low-resource settings. This versatility becomes more pronounced when pre-training utilizes multilingual corpora, with fine-tuned models achieving superior accuracy in comparison to monolingual counterparts \cite{xls-r,xlsr,javed}.

A likely contributor to this improved performance is the acquisition of language related knowledge as an auxiliary consequence of pre-training. Previous work found that embeddings learned by SSL training can be grouped by both language class and linguistic genealogy~\cite{xlsr,javed, fan}. In related work, Tjandra \textit{et al$.$} saw that pre-trained model layers varied in their influence on fine-tuning for a language identification (LangID) task~\cite{tjandra}. Over an ablation study, they were able to remove half the original encoding layers from an XLSR \cite{xlsr} model while retaining fine-tuning results competitive with a full model. Interestingly, these results contend against previous arguments that the upper layers of SSL models were related to LangID~\cite{bertphone}.

While these prior findings demonstrate a relation between SSL and language classification tasks, there are caveats. In general, clustering has only been observed with languages and data seen during pre-training. Though it was observed that intermediate layers of SSL models encode varying linguistically relevant information \cite{pasad}, it is understudied how this relates to LangID. Indeed, we know of only one relevant work that found layer-wise correlation between SSL and LangID, but utilized architecture and pre-training specifically targeted for identification tasks~\cite{bertphone}. Further, it remains an open question how such findings can improve performance.

To address these issues, we extract embeddings from intermediate layers of pre-trained models for LangID, evaluating classification accuracy against layer of extraction, and language composition of pre-training data. Our results lead to the following contributions:

\begin{itemize}[noitemsep]
\item We present empirical findings that pre-trained embeddings alone encode sufficient information for LangID, even for unseen domains and languages.
\item We identify optimal LangID encodings in the lower layers of pre-trained models, suggesting a relation between LangID and pre-training subtasks. 
\item We train a model with near state-of-the-art performance on the VoxLingua107 dataset, despite utilizing one-fifth the parameters of high-performing SOTA systems.
\end{itemize}

\section{Dataset and Model Architecture}
\label{sec:method} 

\subsection{Dataset}
Experiments are performed on the VoxLingua107 dataset, a collection of audio recordings scraped from YouTube comprising 6628 hours of audio \cite{voxlingua107}. It contains a training set with 107 languages and an evaluation set of only 33 languages.\footnote{For more details regarding training and evaluation splits, see http://bark.phon.ioc.ee/voxlingua107/} For our experiments, we segment the training set into segments of 4, 5, and 6 seconds. To avoid overfitting on the evaluation set, we stratified shuffle split 10\% of the training set to create a validation set. For all experiments, we report error rate with the validation set and reserve the unmodified evaluation set for benchmark results. 

\subsection{Model}
We use the encoder-decoder framework for LangID. For our encoder, we use pre-trained Conformer\cite{conformer} models of 18 Conformer layers, 8 attention heads, and hidden depth of 512. Decoding is performed using an x-vector paradigm \cite{xvectors}. Outputs from the encoder are pooled by concatenating the average and standard deviation of embedding parameters before projection to a bottleneck vector of length 256. This is followed by a second projection layer to the number of language classes to learn classification through cross-entropy loss. For feature encoding, we use 80-dimension log-mel filter banks with a 25ms window and a stride of 10ms.

\section{Experiments}
\label{sec:experiments}

\subsection{Pre-training}
\label{ssec:pre-training}
For pre-training, we adopt the contrastive loss objective as seen in \cite{wav2vec}. Our primary model (MultiVP) is pre-trained on the VoxPopuli dataset \cite{voxpopuli} for 400k updates with batchsize 2048. Learning rate was annealed using Noam scheduling with a peak learning rate of $1.4\text{e-}3$ and 25k warmup steps. To observe whether behavior is unique to multilingual models or transfers across SSL configurations, we also pre-train a monolingual model (EngLL) on the LibriLight dataset \cite{librilight} with the same hyper-parameters, except with peak learning rate of $2\text{e-}3$. A randomly initialized conformer model of the same size is used for baseline comparisons (Scratch).

\begin{figure}[t]
  {\includegraphics[width=3.35in,height=2.2in]{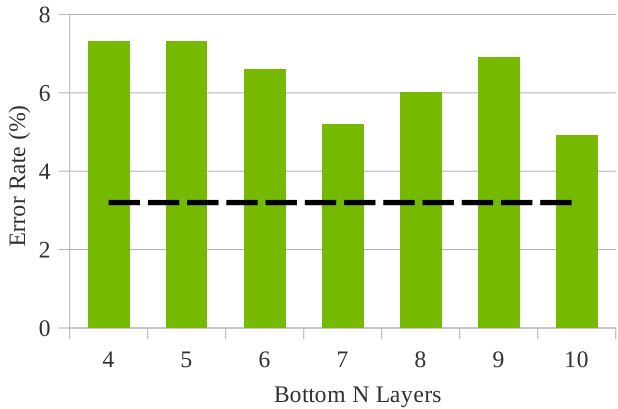}}
    \caption{Best validation error over number of pre-trained layers taken from MultiVP for fine-tuning on VoxLingua107. Layers are relative from the 'bottom' (closest to raw audio) upwards. Baseline is best observed error rate from fine-tuning the all pre-trained layers. Sub-models of 7 and 10 layers outperform other models.}
\end{figure}

\subsection{Fine-tuning}
We begin with a feasibility assessment of fine-tuning pre-trained layers for LangID. Layers are iteratively included from the 'bottom' (layer closest to raw audio) upwards of MultiVP before the resulting sub-model is fine-tuned. For initial exploration, we fine-tune only with sub-models formed of the encoder's 4-10 bottom layers, using the entire model for baseline observations. 

Models are trained over 100 epochs using AdamW optimization with Noam annealing with a warm-up ratio of 0.2. We use SpecAugment for time-masking with steps of 24 width and ratio of .5. Frequency masking is also applied with 4 patches of width 10. We present results from initial fine-tuning ablations with sub-models in \textbf{Fig. 1}, displaying best observed error rates.

While fine-tuning the full model achieved best performance with a validation error rate of 3.2\%, sub-models utilizing only the lower layers still achieved strong results. Further, sub-models of layer sizes 7 and 10 outperformed other sub-models, with error rates of 5.16\% and 4.86\%, respectively. On a review of training behavior, we found these sub-models to have also reached convergence faster than other sub-models and the baseline. We hypothesize this indicates an optimal relationship between certain pre-trained layers and LangID. By exploiting this relationship, we theorize, we can fine-tune models with similar LangID performance, but with substantially fewer parameters.

\subsection{Frozen Encoder}

\begin{table}[t]
\caption{Our 'seen' partition of VoxLingua107, comprising audio from languages present in pre-training for our multilingual model (MultiVP). Reported in hours.}
\begin{tabular}[]{|l|r|r||l|r|r|}
  \hline
  Lang. & Train & Val. & Lang. & Train & Val. \\
  \hline
        Bulgarian & 45.0 & 5 & Italian & 45.9 & 5.1 \\ \hline
        Croatian & 106.2 & 11.8 & Latvian & 37.8 & 4.2 \\ \hline
        Czech & 60.3 & 6.7 & Lithuanian & 73.8 & 8.2 \\ \hline
        Danish & 25.2 & 2.8 & Maltese & 59.4 & 6.6  \\ \hline
        Dutch & 36.0 & 4.0 & Polish & 72.0 & 8.0 \\ \hline
        English & 44.1 & 4.9 & Portuguese & 57.6 & 6.4 \\ \hline
        Estonian & 34.2 & 3.8 & Romanian & 58.5 & 6.5 \\ \hline 
        Finnish & 29.7 & 3.3 & Slovak & 36.0 & 4.0 \\ \hline 
        French & 60.3 & 6.7 & Slovene & 108.9 & 12.1 \\ \hline
        German & 35.1 & 3.9 & Spanish & 35.1 & 3.9 \\ \hline 
        Greek & 59.4 & 6.6 & Swedish & 30.6 & 3.4 \\ \hline
        Hungarian & 65.7 & 7.3 & \textbf{Total}  & \textbf{1216.8} & \textbf{135.2} \\ \hline
\end{tabular}
\end{table}

\begin{table}[t]
\centering
\caption{Our 'unseen' partition of VoxLingua107, comprising audio from languages not present in pre-training for our mutilingual model (MultiVP). Reported in hours.}
\begin{tabular}[]{|l|r|r||l|r|r|}
  \hline
  Lang. & Train & Val. & Lang. & Train. & Val. \\
  \hline
        Assamese & 138.5 & 15.5 & Japanese & 50.4 & 5.6 \\ \hline
        Bengali & 49.5 & 5.5 & Korean & 69.3 & 7.7 \\ \hline
        Basque & 26.1 & 2.9 & Lao & 37.8 & 4.2 \\ \hline
        Hindi & 86.4 & 9.6 & Thai & 54.9 & 6.1 \\ \hline
        Urdu & 37.8 & 4.2 & Mongolian & 63.9 & 7.1 \\ \hline
        Indonesian & 36 & 4.0 & Turkish & 53.1 & 5.9 \\ \hline
        Malay & 74.7 & 8.3 &  Turkmen & 76.5 & 8.5 \\ \hline
        \textbf{Total} & \textbf{855.9} & \textbf{95.1} &  - & - & -\\ \hline
\end{tabular}
\end{table}

To evaluate this hypothesis, we repeat the previous experiment but freeze the pre-trained layers to prevent training during fine-tuning. Given the (relative) simplicity of our decoder, we reason, this would limit performance on LangID to behavior from pre-training. Further, to observe whether results are encoding class information from pre-training or information pertinent to LangID in general, we partition the data into three subsets:

\begin{itemize}[noitemsep]
\item 'seen' contains solely the 23 languages from VoxPopuli observed during pre-training for (MultiVP).
\item 'unseen' contains 14 languages that are largely genealogically distinct from pre-training languages. In our selection, we choose languages with high similarity to (i.e. are a dialect of or mutually intelligible with) at least one other language within the partition to add difficulty to the classification task.
\item '-VP' contains all languages in VoxLingua107 disjoint with 'seen'.
\end{itemize}
Training languages and hours of 'seen' and 'unseen' subsets are provided in \textbf{Table 1} and \textbf{Table 2}, respectively. 

We train and evaluate MultiVP on both 'seen' and 'unseen' for sub-models comprised of layers up to 1-18. For EngLL and Scratch, we only use the 'seen' subset, as we see no theoretical distinction between subsets for monolingual/randomized models beyond the range of possible labels. After obtaining results on these partitions, we repeat the frozen ablation studies with MultiVP on the full VoxLingua107 dataset and '-VP'. Rather than using all sub-models, we use only those of up to layers 6-10, as they consistently showed high performance on previous experiments while still reducing the number of training parameters.

\begin{figure}[t]
{\includegraphics[width=3.42in,height=2.15in]{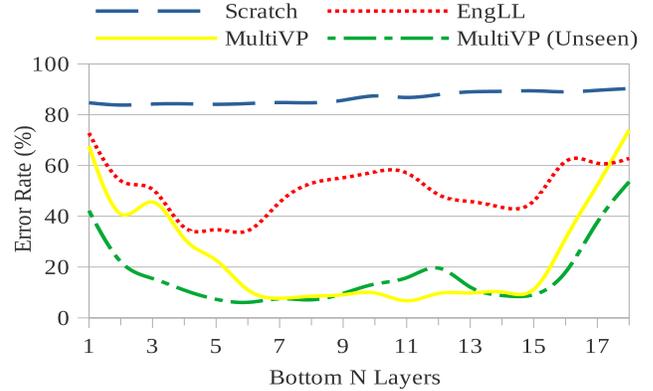}}
\caption{Best validation error over number of pre-trained layers taken from MultiVP, EngLL, and Scratch for fine-tuning over VoxLingua107 partitions. Layers are relative from the 'bottom' of pre-trained models (closest to raw audio) upwards and remain frozen during training. All sub-models are trained over 'seen' and only MultiVP sub-models are trained over 'unseen'. EngLL and MultiVP sub-models of lower-third and middle layers outperform other models.}
\end{figure}

\begin{figure}[h!]
{\includegraphics[width=3.42in,height=2.15in]{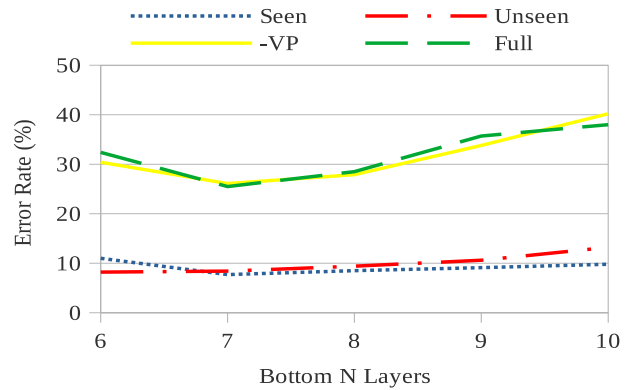}}
\caption{Best validation error over number of pre-trained layers taken from MultiVP for fine-tuning over VoxLingua107 and 'seen', 'unseen', and '-VP' partitions. Layers are relative from the 'bottom' of pre-trained models (closest to raw audio) upwards and remain frozen during training.}
\end{figure}

Results for Scratch, MultiVP, and EngLL on the 'seen' partition and  MultiVP on the 'unseen' partition are displayed in \textbf{Fig. 2}. Pre-trained sub-models consistently perform above chance, with MultiVP having higher results. Further, we observe best performance to generally occur with sub-models formed of layers up to those between the lower third and bottom half of pre-trained models. Improved performance is also achieved using layers up to between the middle and upper third layers, being particularly pronounced when language classes are disjoint with pre-training data. Scratch performance remained almost invariant, indicating architecture alone to have little effect on this behavior. 

Interestingly, MultiVP sub-model performance on the 'unseen' dataset is relatively high. While this can be due to the smaller class size, we note that the chosen languages bear high mutual intelligibility between each other (e.g. Hindi and Urdu, Malay and Indonesian, Thai and Lao) and should cause miscategorization between classes. Further, that performance remained significantly above chance even for identification of the 84 unseen languages in '-VP' (\textbf{Fig. 3}) suggests strong language discrimination for unknown classes.

\subsection{Benchmarking}
\label{ssec:finetune_results}
Given the recurring high performance in the observed layers, we hypothesize that pre-training had distributed language discriminatory information across specific regions of the model (e.g. 'critical regions'). The recurring high performance indicates these layers belong to one of these regions and are ideal for LangID training. To explore this hypothesis, we reattempt fine-tuning with these layers. We use the same configuration as our first experiment but limit training to only 20 epochs to avoid possible overfitting. Further, we attempt to improve performance with augmentation from the Room Impulse Response (RIR) and Noise \cite{RIR} corpora and speed perturbation with 0.95x and 1.05x speeds. As well, we average the five best checkpoints with respect to validation loss. Results and comparisons with other models are displayed in \textbf{Table 3}. 

Despite being only half the size of our original model, the nine-layer sub-model significantly outperforms when evaluated on the VoxLingua107 evaluation set. Further, its error rate surpasses the original XLS-R results \cite{xls-r}. Seen in \textbf{Table 4}, this is a result of improved performance across both short (0...5s) and long (5...20s) utterances.\footnote{As XLS-R-attentive is not publicly available, we are unable to also compare its performance by utterance length.}
While XLS-R-attentive\cite{xlsr_accented} reports better accuracy, we note that our model is within 15\% relative error rate, despite possessing only one-fifth the parameters.

\begin{table}[t]
\caption{Best evaluation error by layer of embedding extraction from MultiVP for fine-tuning over VoxLingua107, with XLS-R baselines. Conformer sub-models use RIR impulse corpora and average five best checkpoints with respect to validation loss. Best model results are in \textbf{bold} while best Conformer results are \underline{underlined}.}
\begin{tabular}{| m{9em} | m{2.4em}  m{3.5em}  m{3.5em} |}\hline
Architecture & Layers & Size, M & \% Error \\ \hline
Conformer & 7 & 52.2 & 7.27 \\ 
 & 8 & 58.5 & 6.90  \\ 
 & \underline{9} & \underline{64.8} & \underline{5.41}  \\ 
 & 10 & 71.1 & 6.46  \\ 
 & 18 & 121 & 6.71  \\ 
\hline
XLS-R \cite{xls-r} & - & 300  & 5.7 \\ 
\textbf{XLS-R-attentive} \cite{xlsr_accented} & - & \textbf{300} & \textbf{4.7}  \\ 
\hline
\end{tabular}
\end{table}

\begin{table}[]
\caption{Comparison of evaluation error rates for best performing Conformer model and XLS-R by length of utterance. Conformer sub-model uses bottom 9 layers from MultiVP as encoder. Fine-tuning performed with RIR impulse corpora and averages five best checkpoints with respect to validation loss. Best model results for each utterance length and average are in \textbf{bold}.}
\begin{tabular}{| m{9em} | m{2.4em}  m{3.5em}  m{3.5em} |}\hline
Model & 0...5s & 5...20s  & avg \\ \hline
 Conformer (L=9) & \textbf{8.7} & \textbf{4.79} & \textbf{5.41}  \\ 
XLS-R & 9.1 & 5.0  & 5.7 \\
\hline
\end{tabular}
\end{table}

\section{Discussion}
\label{sec:discussion}
Our results reinforce the findings of Tjandra \textit{et al.} \cite{tjandra}: when applied to LangID, pre-trained models can maintain competitive performance with only a few bottom layers of the original SSL model used for fine-tuning. Indeed, we discover layer selection to be critical for LangID. 

From the results of our "frozen" ablation studies, we find that a pre-trained model can perform LangID well even without fine-tuning; i.e. a model trained with contrastive loss encodes sufficient language discriminant information into contextual embeddings. Training a simple statistical pooling decoder can produce robust performance over both unknown languages and speech conditions. Further, LangID performance peaks at similar layers of pre-trained models across various training configurations.

While the higher accuracy on observed languages could entail an overlap between LangID and contrastive loss pre-training, the robust performance on unseen languages suggests a more broad rationale. We believe instead that the observed 'critical regions' are locations where speech sounds (phones) become most distinct during pre-training. In related work, Pasad \textit{et al$.$} \cite{pasad} saw medial layers of wav2vec models exhibit high accuracy in phone identification. Their results roughly correlate with our observations over which layers are optimal for LangID. We note that ranking n-gram frequency of discrete segments is one of the classical methods for LangID \cite{cavnar_trenkle} and this task can be accomplished by a single feed-forward layer \cite{shen_joshi}. From this point of view, LangID for speech could be seen as functionally equivalent to learning a language's phonemic distribution. 

\section{Conclusion}
In this paper, we explore using self-supervised learning models for spoken language identification using Conformer architecture. We find that, (1) Pre-trained models optimally encode language discriminating information in lower layers, (2) Embeddings extracted from these layers are robust to unseen languages and acoustic conditions, making them effective candidates for zero-shot language identification, and (3) Training a language identification model initialized with lower layers of a pre-trained model produced results on the VoxLingua107 dataset close to state-of-the-art systems with only a fraction of parameters. 

We open-source the highest performing benchmark model through the NVIDIA NeMo toolkit \cite{nemo}.

\bibliographystyle{IEEEbib}
\bibliography{refs}
\vfill\pagebreak
\end{document}